\documentstyle[aps,pre,multicol,epsf]{revtex}
\def\bea{\begin{eqnarray}}
\def\eea{\end{eqnarray}}

\begin{document}
\title{Coarsening and Pinning in the Self-consistent Solution of Polymer
Blends Phase-Separation Kinetics}
\author{Claudio Castellano$^a$ and Federico Corberi$^b$}
\address{$^a$Istituto Nazionale di Fisica della Materia, Unit\`a di Napoli
and Dipartimento di Scienze Fisiche, Universit\`a di Napoli,
Mostra d'Oltremare, Pad. 19, 80125 Napoli, Italy}
\address{$^b$Dipartimento di Fisica, Universit\`a di Salerno, 84081
Baronissi(SA), Italy}
\maketitle
{\abstract We study analytically a continuum model for phase-separation
in binary polymer blends based on the Flory-Huggins-De Gennes free 
energy, by means of the self-consistent large-$n$ limit approach.
The model is solved for values of the parameters corresponding to the
weak and strong segregation limits. For deep quenches
we identify a complex structure of intermediate regimes and crossovers
characterized by the existence of a time domain such that
phase separation is pinned, followed by a preasymptotic
regime which in the scalar case corresponds to surface diffusion. 
The duration of the pinning is analytically computed and diverges
in the strong segregation limit.
Eventually a late stage dynamics sets in, described by scaling
laws and exponents analogous to those of the corresponding small molecule
systems.}

\section{Introduction}
The kinetics of phase separation has been the subject of considerable
effort in recent years~\cite{Gunton83}.
Among the most investigated systems are binary polymer
blends~\cite{Binder94};
beyond their
importance for technological applications, such systems are
extremely interesting also from a more fundamental point of view.
Due to the intrinsic connectivity of macromolecules, phase separation
in polymeric mixtures takes place on space and time scales 
more easily accessible experimentally than for small molecule systems.
Furthermore, long range interactions along the chains greatly reduce
the size of the critical region, allowing to disregard critical
fluctuations in many situations.
On the other hand, connectivity gives rise to additional complexity
in the system, that can in principle lead to a different behavior
with respect to the small molecule case.
Previous investigations~\cite{Chakrabarti89} have shown that the
global theoretical picture is the same for the asymptotic dynamics
of a system in the weak segregation limit (WSL),
but still many challenging problems remain unsolved:
for example the origin of the pinning phenomenon observed in
experiments for off-critical quenches and the theoretical
study of the strong
segregation limit (SSL), which is hard to attack with the traditional
numerical methods.

In this paper we consider the most common theoretical model describing
the kinetics of phase separation for a binary polymer blend:
the Cahn-Hilliard equation with the Flory-Huggins-De Gennes (FHDG)
free energy functional.
We propose a generalization of the above mentioned model to the case
of an $O(n)$ vector order parameter field $\psi_{\alpha}({\bf x},t)$,
with $\alpha=1,...,n$, and study  the resulting equations in the
large-$n$ limit.
The extrapolation of the model to large-$n$ is a widely used technique
of statistical mechanics, well suited for Ginzburg-Landau-type models,
which has revealed quite powerful in the study of several
systems~\cite{Coniglio94}.
The large-$n$ limit allows to deal with the nonlinearities
of the model by means of a self-consistency prescription which
effectively linearizes the equations.
The solution of the large-$n$ model is often a good approximation
to the evolution of physical systems with finite $n$. 

The application of this method to the equation of motion for a
symmetric polymer blend allows one to write down closed form equations
for the main observables and to study analytically the time evolution
of the model for all values of the parameters.
In this way we can present results not only for the often investigated
WSL but also for the much less known SSL,
where a complex preasymptotic scenario can be identified.

In particular we find that the very late evolution belongs
to the same  universality class of the small molecule case and
the typical domain size grows as $t^{1/z}$ with the
dynamical exponent $z=4$, as usual in vectorial systems. 
For intermediate times, instead, different paths can be taken.
More precisely, while  for shallow quenches (WSL) the evolution
closely reproduces what is known for small molecules,
for deep quenches (SSL) additional terms, introduced in the equation
of motion by the polymeric nature of the system, can play a relevant
role.
In this case, after the linear Cahn-Hilliard instability has occurred,
the dynamics becomes exceedingly slow and the non equilibrium blend
remains in a pinned state over an appreciable time interval.
This effect is due to the existence of a nonequilibrium
partially-ordered state which becomes stable for infinitely deep
quenches or infinitely long chains.
For quenches of finite depth the system remains close to this state for a
characteristic time $\tau_p$ which is shown to diverge in the SSL
as a power law or exponentially as the limits of the parameters
are taken in different order.
Later on it enters a second preasymptotic regime characterized by
a slow coarsening of domains whose typical size grows as $t^{1/6}$.
A crossover leads finally to the late stage where domains
grow as $t^{1/4}$.

These results are derived using the large-$n$ model for phase-ordering
in polymer blends, which is an approximation of the real scalar model.
It has been recognized in the small molecule case that this
approximation leads to some results different
from the properties of the corresponding scalar systems.
One of them is trivial: the exponents found using such method are valid
for vectorial systems and therefore differ from those valid for scalar
systems: the exponent $z=4$ for the growth of domains 
in the large-$n$ model is known to correspond
to $z=3$ when the order parameter is scalar.
Another known discrepancy is that within the large-$n$ approximation
the structure factor obeys the multiscaling symmetry while systems for
finite $n$ exhibit scaling symmetry.
This is related to the different nature of the phase-transition
when $n$ is infinite and in particular to the properties of the 
equilibrium state the large-$n$ model evolves towards;
although displaying the correct structure factor such state is not truly
ordered: the local order parameter distribution remains
gaussian instead of becoming bimodal as effect of the formation of
domains \cite{Castellano97}.
Therefore a multiscaling structure factor in the large-$n$ model
must be interpreted as reflecting scaling symmetry in real systems.
These differences are well known from the small molecule case and
can be easily taken into account,
but the absence of topological defects in the large-$n$ model may
make meaningless the extrapolation of the results for
the large-$n$ model to the description of actual evolution in
scalar systems.
More precisely one may wonder whether a model with no interfaces
(the large-$n$ model) can describe with sufficient accuracy the
dynamics in presence of a field-dependent mobility,
distinguishing bulk (where $|\psi| \simeq 1$) from interfaces
(where $|\psi| \simeq 0$).
This problem is discussed at length throughout the paper.
The key point for the answer is a close comparison of our equations
with those of the large-$n$ model for ordinary Ginzburg-Landau systems.
In such systems the connsection between large-$n$ and scalar results is
well established.
This allows the interpretation of all the complex pattern of regimes
and crossovers found within the large-$n$ model for polymeric systems
in terms of concepts like bulk and surface diffusion even if they cannot
be defined when $n=\infty$.
Hence it turns out that the analytical approximate solution preserves
most of the essential physics of phase-ordering also for systems more
complex than small molecule mixtures.

The paper is organized as follows.
In Section~\ref{Sec2} the usual equation describing phase separation for 
polymers
is introduced and the large-$n$ limit model is deduced from it.
Section~\ref{Sec3} is devoted to the analytical solution of this model and
to the identification of the different regimes.
The findings of Section~\ref{Sec3} are verified and completed by means of a 
numerical solution of the large-$n$ limit equations in Section~\ref{Sec4}.
Finally in Section~\ref{Sec5} the results are discussed and conclusions are
drawn.

\section{The model}
\label{Sec2}

Model blends are typically described by the Flory-Huggins-De Gennes free
energy functional \cite{DeGennes77,Pincus81}
\begin{equation}
\frac{F(\psi )}{k_BT}=\int d{\bf x} 
\left [ \frac{f_{FH}(\psi )}{k_BT}+
\kappa( \psi)
|\nabla  \psi ({\bf x},t)|^2 \right ]
\label{ham}
\end{equation}
with
\begin{equation}
\kappa( \psi)=\frac{\sigma_A^2}{18[1+\psi ({\bf x},t)]}+
\frac{\sigma_B^2}{18[1-\psi ({\bf x},t)]}+\chi\lambda ^2
\label{kappa}
\end{equation}
where $\psi ({\bf x}, t)$ is the order parameter field, 
$k_B$ is the Boltzmann constant,
$\sigma _A$
and $\sigma _B$ are the Kuhn lengths of the two species,
$\lambda$ is an effective interaction distance between monomers, 
and the Flory-Huggins free energy is \cite{Flory41}
\bea \nonumber
\frac{f_{FH}(\psi)}{k_BT} &= &\frac{1+\psi ({\bf x},t)}{2N_A}\ln [1+\psi
({\bf x},t)]+
\frac{1-\psi ({\bf x},t)}{2N_B}\ln [1-\psi ({\bf x},t)]+ \\
& & +\frac{\chi}{4} [1-\psi^2 ({\bf x},t)]
\label{localpot}
\eea
where $N_A$ and $N_B$ are the degrees of polymerization of chains
A and B; $\chi$ measures the strength of repulsion between unlike
molecules and is inversely proportional to temperature.
In the following we will always consider
for simplicity a symmetric blend, for which $N_A=N_B=N$ and
$\sigma_A=\sigma_B=\sigma$.
In this case the critical value $\chi N = 2$ separates stable states of
the blend ($\chi N<2$) from the thermodynamically unstable region ($\chi N>2$)
where the mixture decays in two separate phases.
The order parameter is related to the volume fraction of
A-molecules by $\psi ({\bf x},t) = 2 \phi ({\bf x},t)-1$.

The theoretical description of spinodal decomposition in binary blends is
based on the Cahn-Hilliard equation for the time evolution of the order
parameter field, originally introduced for small molecule 
systems \cite{Cahn58}
\begin{equation}
\frac {\partial \psi ({\bf x},t)}{\partial t}={\bf \nabla} \cdot \left [
M(\psi ){\bf \nabla} \frac{\delta F(\psi)}{\delta \psi}\right ]
+\eta({\bf x},t)
\end{equation}
where $M(\psi)$ is the mobility and $\eta({\bf x},t)$ is a gaussian
white noise with zero average and variance proportional to the
temperature.
For a symmetric blend the mobility
\begin{equation}
 M(\psi)=\frac{ND}{4} [1-\psi^2 ({\bf x},t)]
\label{mobility}
\end{equation}
has been proposed \cite{DeGennes77}
where $D$ is a self-diffusion coefficient. 
With these positions the Langevin evolution equation of the order
parameter field is given by
\bea \nonumber
\frac{\partial \psi ({\bf x},t)}{\partial t} & =& 
\frac{ND}{2}{\bf \nabla}\cdot
\left \{[1-\psi ({\bf x},t)^2]{\bf \nabla}\left [\frac{1}{4N}\ln \left( \frac 
{1+\psi ({\bf x},t)}{1-\psi ({\bf x},t)}\right) -{\chi \over 4} 
\psi ({\bf x},t) +
\right. \right. \\ \nonumber
& & -\left(\lambda ^2 \chi+\frac {\sigma ^2}{9[1-\psi ({\bf x},t)^2]}
\right ){\bf \nabla} ^2 \psi ({\bf x},t) + \\ 
& & \left. \left. -\frac{\sigma ^2\psi ({\bf x},t)}
{9[1-\psi ({\bf x},t) ^2]^2}
[{\bf \nabla} \psi ({\bf x},t) ]^2\right ] \right \}
\label{n=1}
\eea
where we have neglected the thermal noise term since  it is possible to
show, at least in the analog of this equation for
small molecule systems \cite{Rogers88,Bray89}, that the temperature
is asymptotically an irrelevant parameter below the order-disorder line.
The scalar model introduced so far can be generalized to the case of
an $n$-component vectorial order parameter $\psi({\bf x},t) \equiv
\{ \psi_{\alpha}({\bf x},t)\}$ with $\alpha = 1, \dots, n$.
As described in Appendix 1, when the number $n$ of components diverges
the equation of motion for the single component $\psi_{\alpha}({\bf x},t)$
(we drop the index $\alpha$ in the following) reads
\bea \nonumber
\frac{\partial \psi ({\bf k},t)}{\partial t}&=&-\frac{ND}{2}[1-S(t)]k^2
\left \{\frac{1}{4NS^{\frac{1}{2}}(t)}\ln \left( 
 \frac 
{1+S^{\frac{1}{2}}(t)}{1-S^{\frac{1}{2}}(t)} \right)
-{\chi \over 4} + \right. \\
& & \left. 
+\frac{\sigma ^2}{9}\frac{S_2(t)}{[1-S(t)]^2}
+\left (\lambda ^2
\chi+\frac {\sigma ^2}{9[1-S(t)]}\right )k^2 
\right \}\psi ({\bf k},t)
\label{n=oo}
\eea
where by definition $S(t) \equiv <\psi^2({\bf x},t)>$ and
$S_2(t) \equiv <\left [\nabla \psi({\bf x},t)\right]^2>$.

In the following the dynamics of systems undergoing critical quenches
($<\psi ({\bf x},t)>=0$) will be studied.
In this case the quantities $S(t)$ and $S_2(t)$ can be computed
self-consistently
using the structure factor
$C({\bf k},t)=<\psi ({\bf k},t)\psi(-{\bf k},t)>$, the Fourier
transform of the real space pair connected correlation function, through
\begin{equation}
S(t)=\int _{|{\bf k}|<q}\frac{d{\bf k}}{(2\pi)^d}C({\bf k},t)
\label{eqS}
\end{equation}
and
\begin{equation}
S_2(t)=\int _{|{\bf k}|<q}\frac{d{\bf k}}{(2\pi)^d}k^2C({\bf k},t)
\label{eqS2}
\end{equation}
where $d$ is the spatial dimension of the system and $q$ is a
phenomenological ultraviolet momentum cutoff.
The evolution equation of the structure factor can be obtained from
Eq.~(\ref{n=oo}) as
\bea \nonumber
\frac{\partial C ({\bf k},t)}{\partial t}&=&-ND[1-S(t)]k^2
\left \{\frac{1}{4NS^{\frac{1}{2}}(t)} \ln \left[ 
 \frac {1+S^{\frac{1}{2}}(t)}{1-S^{\frac{1}{2}}(t)}
\right] - {\chi \over 4} + \right. \\
& & \left. +\frac{\sigma ^2}{9}\frac{S_2(t)}{[1-S(t)]^2}
+\left (\lambda ^2 \chi+\frac {\sigma ^2}{9[1-S(t)]}
\right )k^2 
\right \}C({\bf k},t)
\label{eqC}
\eea
Eq.~(\ref{eqC}) together with the self-consistency relations (\ref{eqS})
and (\ref{eqS2}) constitute the integro-differential equations
governing the dynamics of the FHDG model in the large-$n$ limit,
and will be the object of our study.
Some differences occur between Eq.~(\ref{eqC}) and its analog for
small molecule systems{\cite{Coniglio94}:
the overall time-dependent factor $1-S(t)$ reflects the
presence of a field-dependent mobility as opposed to the constant
value usually taken;
the terms proportional to $\sigma^2$ are a consequence of the order
parameter dependence of $\kappa(\psi)$ and do not appear in the
equation for small molecules;
the logarithmic form of $f_{FH}$ is different from the
usual Ginzburg-Landau quartic potential.

As we will see, in phase-separation of binary polymer blends a key role
is played by the product $\chi N$, setting how strongly segregated the two
species are.
When $\chi N \gtrsim 2$ the absolute value of the equilibrium order
parameter in the coexisting phases is much smaller than 1, indicating
that in A-rich regions
a high concentration of B molecules is present and vice versa.
This is the Weak Segregation Limit (WSL), which is also called shallow
quench condition, since it is usually realized by lowering $T$ just below
$T_c$.
For deep quenches instead $\chi N \gg 2$ (Strong Segregation Limit, SSL);
the order parameter equilibrium value is very close to $\pm 1$ because
separated domains are almost pure. 

\section{Analytical study of the model}
\label{Sec3}

In the following a high temperature disordered initial condition
$C({\bf k},0)=\Delta$ will be considered.
Upon introducing the three quantities
\begin{equation}
L(t)=\left(\frac{1}{9}ND\sigma ^2 t\right)^{\frac{1}{4}}
\label{eqL}
\end{equation}
\begin{equation}
\Lambda (t)=\left \{ND\lambda ^2\chi \int _0 ^t 
[1-S(\tau)]d\tau \right \}^{\frac{1}{4}}
\label{eqLambda}
\end{equation}
and
\bea \nonumber
{\cal L}^2(t)&=&-ND\int _0 ^t \left \{ {[1-S(\tau)]\over 4}\left
[\frac{1}{NS^{\frac{1} {2}}(\tau)}
\ln \left [\frac {1+S^{\frac{1}{2}}(\tau)}{1-S^{\frac{1}{2}}(\tau)} \right ]
-\chi \right ] + \right. \\
& & \left. +\frac{\sigma ^2}{9}\frac{S_2(\tau)}{1-S(\tau)}\right \}
d\tau
\label{eqCalL}
\eea
Eq.~(\ref{eqC}) can be formally integrated, yielding
\begin{equation}
C({\bf k},t) =\Delta \exp \left\{k^2\left[\Lambda^4(t)+L^4(t)\right]
(2 k_m^2(t)-k^2)\right\} 
\label{eqInt}
\end{equation}
where
\begin{equation}
k^2_m(t)=\frac{{\cal L}^2(t)}{2[L^4(t)+\Lambda ^4(t)]}
\label{kmax}
\end{equation}
is the position of the peak provided ${\cal L}^2 >0$, as is the case for
sufficiently long times in the phase-ordering region.

The introduction of the three quantities~(\ref{eqL}), (\ref{eqLambda}) and
(\ref{eqCalL}) allows the description of the
dynamical evolution in terms of their competition. Depending on
their relative size, the system exhibits different properties in
subsequent time regimes, as will be described in detail below.

\subsection{Early stage}

For short times after the quench, assuming the initial fluctuations $\Delta$
are not large, $S(t)$ and $S_2(t)$ can be neglected in Eq.~(\ref{eqC})
and the system exhibits the usual linear behavior of phase-ordering.
The time evolution of $\Lambda(t)$ and ${\cal L}(t)$ is easily computed
\begin{equation}
\Lambda (t)=(N D \lambda ^2 \chi t)^{\frac{1}{4}} \sim L(t)
\label{LamCrit}
\end{equation}
and
\begin{equation}
{\cal L}(t)=\left[-{N D \over 4}
\int_0^t d\tau (-\chi+2/N)\right]^{\frac{1}{2}}
= \left[{D\over 4}(\chi N-2) t \right]^{\frac{1}{2}}
\end{equation}
As a consequence, in complete analogy with linear behavior in small
molecule systems, the position of the maximum remains constant
\begin{equation}
k_m(t) = \left[{(\chi N -2) \over 8N(\chi \lambda^2 + \sigma^2/9)}
\right]^{\frac{1}{2}} \equiv k_0
\end{equation}
while its height grows exponentially fast
\begin{equation}
C(k_m,t) = \Delta \exp\left[{D(\chi N-2)^2 \over 64N(\chi \lambda^2+
\sigma^2/9)}t\right]
\end{equation}
During the very early stages of this linear regime
(for times $t \ll t^*$, where $t^*$ will be determined below)
the structure factor obeys an approximate scaling form.
This can be shown by considering that, according to Eq.~(\ref{eqInt}),
$C({\bf k},t)$ decays for $|{\bf k}| > k_m$ over the typical distance
$W(t) =
[\Lambda^4(t)+L^4(t)]^{-1/4} = [ND(\chi \lambda^2 + \sigma^2/9)t]^{-1/4}$.
Since for small times $W(t)$ is very large, the integrals defining $S(t)$
and $S_2(t)$ are dominated by the contributions for large momenta.
For large $k$, $k_m^2$ can be neglected in Eq.~(\ref{eqInt}) and
the structure factor can be written as
\begin{equation}
C(k,t) = \Delta \exp\left\{-k^4[\Lambda^4(t)+L^4(t)]\right\}
= \Delta \exp\left\{-g [kL(t)]^4 \right\}
\label{earlyscaling}
\end{equation}
with $g$ constant.
Hence $C(k,t)$ exhibits for $k \gg k_m$ a scaling form with respect
to the growing length $L(t)$;
$S(t)$ and $S_2(t)$ can be computed easily, yielding
\begin{equation}
S(t)\sim \Delta L(t)^{-d} \hspace {2 cm}
S_2(t)\sim \Delta L(t)^{-(d+2)}
\label{eqSS}
\end{equation}
consistently with the assumption that they are small.
This very early scaling regime is completely analogous to the regime found
in small molecule systems for very short times before the usual
Cahn-Hilliard linear regime \cite{Corberi95}.
Its physical origin is the presence of totally uncorrelated fluctuations
in the initial state, creating large gradients in concentration between
neighboring regions: the square gradient is the dominating contribution
to the excess free energy.
Then the system lowers its free energy by reducing everywhere the local order
parameter so that the contribution of the square gradient is reduced.
In this way the energy associated with the local potential grows, but as long
as it is much smaller than the other contribution it does not affect the
evolution: the behavior is diffusive as if the system would be at $\chi N=2$.
This type of evolution ends when the global reduction of the local order
parameter
ends up increasing the total free energy, because the local contribution
grows more than the decrease in the square gradient one.
In terms of the structure factor this happens when the peak position is no
longer much smaller than its width $W(t)$, i.e. for $t=t^*$ such that
\begin{equation}
k_m(t^*) \simeq [\Lambda^4(t^*) + L^4(t^*)]^{-1/4}
\end{equation}
from which
\begin{equation}
t^* = {64N (\chi \lambda^2+\sigma^2/9) \over D(\chi N-2)^2}
\label{t_crossover}
\end{equation}
From Eq.~(\ref{t_crossover}) we conclude that the duration of the very
early scaling regime diverges for infinitely shallow quenches.
In the same limit $k_m$ vanishes and therefore the scaling
form~(\ref{earlyscaling}) holds down to $k=0$.
After the crossover time $t^*$ the usual linear behavior sets in and the order
parameter saturates exponentially fast to the local equilibrium, leading to
an exponential growth of $S(t)$ and $S_2(t)$.

\subsection{Pinned regime}
\label{SecPin}

The linear regime continues until a time $t_0$ such that $S(t_0)$
is close to the equilibrium value $S(\infty)$ which corresponds to the
minima of the Flory-Huggins local potential.
The subsequent behavior of the system strongly depends on the value
of $\chi$, $N$ and $\sigma$.
In the WSL $S(\infty) \ll 1$ and therefore $\Lambda(t) = (ND \chi
\lambda^2t)^{1/4 }\sim L(t)$; the system enters immediately the asymptotic
stage described in Sec.~\ref{SecLabel}.
In the SSL instead,
the dynamical evolution of the large-$n$ equations enters a quasi-stationary
regime which extends over a time domain diverging when 
$\chi N \to \infty$ and $\chi / \sigma^2 \to \infty$; this case
will be referred to as the pinning limit. During this time interval no
appreciable evolution of $C({\bf k},t)$ is observed and the blend
is practically pinned in a configuration characterized by 
phase-separated domains of finite size.
This phenomenon reflects the existence of a static non-equilibrium
configuration becoming stable in the pinning limit.
In order to see this let us consider the static solutions of the model.
For general values of the parameters it is clear that
\begin{equation}
C({\bf k},\infty)=(2\pi)^d S_{\infty} \delta ({\bf k})
\label{equi}
\end{equation}
is a static solution ($\partial C/ \partial t = 0$) of Eq.~(\ref{eqC}),
where the choice of $S_{\infty} \equiv S(\infty)$ as prefactor is dictated
by the self-consistency condition~(\ref{eqS}).
The value of $S_{\infty}$ is fixed by requiring the solution~(\ref{equi})
to be a minimum of the FHDG free energy.
The resulting condition is
\begin{equation}
\frac{1}{NS_\infty^{\frac{1}{2}}}\ln \left( \frac{1+S_\infty^{\frac{1}{2}}}
{1-S_\infty^{\frac{1}{2}}}\right ) -\chi =0
\label{statS}
\end{equation}
which in the scalar case indicates that the order parameter lies on
the minima of the local potential.
This is the equilibrium state towards which the system evolves for all finite
values of $\chi N$ and $\chi /\sigma^2$, as can be checked using the asymptotic
results of Sec.~\ref{SecLabel}.
A different situation occurs instead when $\chi N$ and $\chi /\sigma^2$ both
diverge.
In this limit the equation of motion reads
\begin{equation}
\frac{\partial C ({\bf k},t)}{\partial t}=-\chi ND[1-S(t)]k^2
\left \{-{1 \over 4}+\lambda^2 k^2 \right \}C({\bf k},t)
\label{eqp}
\end{equation}
Integrating one obtains for all times
\begin{equation}
C(k,t) = \Delta \exp\left\{-k^2\Lambda^4(t) \left(k^2-{1 \over 4\lambda^2}
\right)\right\}
\end{equation}
Therefore a static solution requires $\Lambda(t=\infty) = \Lambda_{\infty} =
\mbox{const}$, yielding
\begin{equation}
C(k,t) \equiv C_p(k) = \Delta \exp\left\{-k^2\Lambda^4_{\infty}
\left(k^2-{1 \over 4\lambda^2}\right)\right\}
\label{Cp}
\end{equation}
We notice from Eq.~(\ref{eqLambda}) that a constant $\Lambda(t)$ implies
$S(t) =1$ and this sets the value of $\Lambda_{\infty}$ via the
self-consistency condition
\begin{equation}
\Delta \int \frac{d{\bf k}}{(2\pi)^d} C_p(k) = 1
\label{Lambdainfty}
\end{equation}

Hence in the pinning limit the system does not evolve towards the free energy
ground state corresponding to complete order, as revealed by the lack
of the Bragg peak at ${\bf k}=0$ in $C_p(k)$.
Instead a state partially ordered over a typical
length $k_m^{-1} = 2 \sqrt{2} \lambda$ is dynamically generated;
despite having a higher free energy than the equilibrium configuration, the
state~(\ref{Cp},\ref{Lambdainfty}) is strictly asymptotic in the pinning limit.
When $\chi N$ and $\chi/\sigma^2$ are large but finite, the additional terms
in Eq.~(\ref{eqC}) destabilize the pinned state: the system gets trapped
around it for a time domain diverging in the pinning limit.
The duration $\tau_p$ of the interval during which no coarsening practically
occurs can be easily estimated, as reported in Appendix 2.
It turns out that this time strongly depends on how the pinning limit is
approached, i.e. on the order of the limits $\chi N \to \infty$ and
$\chi / \sigma^2 \to \infty$.
If one takes $\chi /\sigma^2 \to \infty$ first
\begin{equation}
\tau_p \sim \exp\left\{a\chi N\right\}
\label{tau_exp}
\end{equation} 
where $a$ is a known constant, while when $\chi N \to \infty$ first
\begin{equation}
\tau_p \sim \left({\chi \over \sigma^2} \right)^{1/2}
\label{tau_power}
\end{equation} 
This twofold behavior reflects the different terms that can
destabilize the pinned state. By comparing Eq.~(\ref{eqC}) and
Eq.~(\ref{eqp}) it turns out that  Eq.~(\ref{eqC}) contains two kinds
of additional terms: the logarithmic contribution and the terms
proportional to $\sigma^2$.
The former is proportional to $1/N$ and hence destroys pinning when
$\chi N$ is finite leading to Eq.~(\ref{tau_exp});
the latter are active when $\sigma$ is nonvanishing and cause the decay
of the pinned state over the characteristic time~(\ref{tau_power}).
 
It is important to stress that the pinning regime occurs when phase
separation has already taken place but is still incomplete and, therefore,
has nothing to do with the usual metastability present during phase
ordering, which decays via nucleation.

\subsection{Surface diffusion regime}

In the SSL, after the end of pinning the structure factor has the
form~(\ref{eqInt}).
During this regime $C(k,t)$ is sharply peaked around $k_m(t)$;
this allows the evaluation of $S(t)$ and $S_2(t)$ by saddle
point technique, yielding
\begin{equation}
S(t)\sim {\cal L}^{-d}(t)e^{\frac{{\cal L}^4(t)}{4[L^4(t)+\Lambda^4(t)]}}
\label{asymptS}
\end{equation}
and
\begin{equation}
S_2(t)\sim{\cal L}^{-2}(t)S(t)
\label{asymptS2}
\end{equation}
Requiring the saturation of $S(t)$ to its equilibrium value $S_\infty$
and neglecting $L(t)$ with respect to $\Lambda(t)$, for sufficiently large
$\chi/over \sigma^2$, one has
\begin{equation}
{\cal L}^d(t) \sim e^{\frac{{\cal L}^4(t)}{4\Lambda^4(t)}}
\label{selft1/6}
\end{equation}
According to its definition~(\ref{eqCalL}), 
to leading order in $S_\infty-S(t)$, ${\cal L}^2(t)$ can be written as
\begin{equation}
{\cal L}^2(t)=-ND\int _0 ^t \left ( b
\frac{[S^{1/2}(\tau)-S^{1/2}_\infty][1-S(\tau)]}{4S^{1/2}
(\tau)}
+\frac{\sigma ^2}{9}\frac{S_2(\tau)}{[1-S(\tau)]^2}\right )d\tau
\label{CalL1/6}
\end{equation}
with $b=2/[N(1-S_\infty)]$.
Then, neglecting for sufficiently small Kuhn length $\sigma $ the term
containing $S_2(t)$ in Eq.~(\ref{CalL1/6}), and letting $S_\infty \simeq 1$,
we obtain
\begin{equation}
{\cal L}^2(t) \sim {\cal L}^2(t_0) + \int _{t_0} ^t
[1-S^{\frac{1}{2}}(\tau)]^2d\tau 
\label{eqCalL1/6}
\end{equation}
where $t_0$ is the crossover time from pinning to this regime.
In the same way, one finds
\begin{equation}
{\Lambda}^4(t)\sim {\Lambda}^4(t_0) + \int _{t_0} ^t
[1-S^{\frac{1}{2}}(\tau)] d\tau 
\label{eqLambda1/6}
\end{equation}
Assuming that ${\cal L}(t)$ and $\Lambda(t)$ diverge with time, we neglect
the constant terms in Eqs.~(\ref{eqCalL1/6}) and (\ref{eqLambda1/6}).
Eqs.~(\ref{selft1/6}),(\ref{eqCalL1/6}) and (\ref{eqLambda1/6}) admit then
the solution
\begin{equation}
\Lambda (t)\sim t^{\frac{1}{6}}\hspace {2 cm}
{\cal L}(t)\sim t^{\frac{1}{6}}(\log t)^{\frac{1}{4}}
\label{t=1/6}
\end{equation}
with $1-S^{1/2}(t)\sim t^{-1/3}$, consistently with the assumption of
diverging ${\cal L}(t)$ and $\Lambda(t)$.
As a consequence, using Eq.(\ref{kmax}),
$k_m(t) \sim t^{-1/6} (\log t)^{1/4}$.

During this regime the dynamics is governed by $\Lambda(t)$ which
dominates over $L(t)$.
With the help of definitions~(\ref{eqL}) and ~(\ref{eqLambda}) it
is clear that during this stage the system behaves as if the square
gradient coefficient $\kappa$ was independent of $\psi$, i.e. as if
$\sigma=0$.
With that condition Eq.~(\ref{eqC}) becomes perfectly analogous to
the equation of motion for the large-$n$ approximation
of a Ginzburg-Landau system with a field dependent mobility
$M(\psi)=1-\psi^2$;
it is easily recognizable that also in this case the analytical solution
yields $k_m(t) \sim t^{-1/6}$.
This similarity with the Ginzburg-Landau system helps understanding the
physical meaning of this regime.
The scalar case for the small molecule Ginzburg-Landau problem with
non constant mobility has been studied by
means of a Lifshitz-Slyozov approach~\cite{Bray95} and numerical
simulations~\cite{Lacasta92}.
The outcome of such investigations was that when $n=1$ the typical
length grows as $t^{1/4}$ and the dominating growth mechanism is 
surface diffusion.
This leads to the conclusion that also for a polymeric mixture the time
regime  with $k_m(t) \sim t^{-1/6}$ in the large-$n$ model reflects
a surface diffusion regime in the corresponding scalar model with
$k_m(t) \sim t^{-1/4}$.
This is why (with an abuse of language) we termed this regime as
surface-diffusive: clearly no surfaces exist in the large-$n$ model
and no diffusion along them takes place.
Nevertheless the behavior of the large-$n$ model clearly reflects the
prevalence of this growth mechanism in the corresponding scalar system.

\subsection{Asymptotic regime}
\label{SecLabel}

The behavior illustrated in the preceding section is not yet asymptotic,
since $\Lambda(t) \sim t^{1/6}$ cannot dominate $L(t) \sim
t^{1/4}$ forever.
When this occurs, the system enters the very late stage of its temporal
evolution.
What follows is valid also for the WSL; in such case $L(t)$ 
and $\Lambda(t)$ are always proportional to $t^{1/4}$ and the
asymptotic regime begins right after the linear one.

The analytic treatment of this late regime is again based on the 
saddle point technique, yielding Eq.~(\ref{asymptS}), and the request
that the order parameter field lies on the ground state manifold,
$S(t) = S_{\infty}$.
From Eq.~(\ref{eqL}) and Eq.~(\ref{eqLambda})
$\Lambda (t)\sim L(t) \sim t^{1/4}$ then
Eq.~(\ref{asymptS}) gives
\begin{equation}
{\cal L}^4(t)\sim L^4(t) \log t
\label{asymptCalL}
\end{equation}
and therefore
\begin{equation}
k_m(t) \sim (\log t/t)^{1/4}
\label{asymptkmax}
\end{equation}

Here the two lengths $L(t)$ and $k_m(t)$ diverge
in the same way up to a logarithmic factor yielding
multiscaling precisely as for 
phase-ordering in large-$n$ ordinary mixtures \cite{Coniglio94} 
and showing
that the two models fall into the same universality class.
Again we can use the comparison with phase-separation in ordinary
mixtures to extrapolate results for scalar polymer blends:
on the basis of what is known \cite{Bray89,Bray92} about small
molecule systems, a scaling regime is expected to be obeyed
also for polymers for finite 
$n$, characterized by a single diverging length growing as 
$t^{1/z}$, with $z=4$ for vector order parameter and $z=3$ in the 
physically relevant case $n=1$.
The coarsening mechanism prevailing during this stage in scalar systems
is bulk diffusion: although strongly suppressed in the SSL, it is not
vanishing and, being associated with a faster domain growth, finally
dominates over surface diffusion.

It is interesting to remark that the model for polymers has the same
asymptotic behavior of small molecules, but for non trivial reasons.
In ordinary mixtures $L(t)$, the dominating length during the late stage,
is formed by the product of a constant mobility times a constant square
gradient coefficient.
For polymer blends $L(t)$ is the result of a non constant $M(\psi)$
times the field dependent part of $\kappa(\psi)$: the expression for $L(t)$
is the same of the Ginzburg-Landau case because the order parameter
dependence in the two factors cancels out.
Therefore, even if the asymptotic behavior of this model for polymer
blends is the same as for small molecules, it is not correct saying
that the field dependence of $M$ and $\kappa$ is irrelevant.

\section{Numerical Results}
\label{Sec4}

In this section we present the results of the numerical solution of
the large-$n$ model.
The solution is performed by simple iteration
of the discretized version of Eqs.~(\ref{n=oo},\ref{eqS},\ref{eqS2})
with $d=3$ and 1024 values of $k$.
The diffusion coefficient $D$ is chosen equal to 4 and
the number of  monomers is fixed to $N=0.25 \times 10^5$.
The value of the parameter $\chi$ is changed over many
orders of magnitude, so that we can clearly distinguish the different
time regimes.

We start by considering the very early stages.
The behavior of $S(t)$ and $S_2(t)$ for very early times is shown
in Fig.~\ref{Fig1}.
The initial decay of $S(t)$ follows
very accurately the power-law $t^{-3/4}$ found analytically.
The same agreement is found for $S_2(t)$ with the decay $t^{-5/4}$.
Notice that the crossover time is very close to the estimate based
on Eq.~(\ref{t_crossover}): $t^* \simeq 4 \times 10^{-2}$.

With regards to the following stages,
in Fig.~\ref{Fig2} $S_2(t)$ and $k_m(t)$ are plotted versus
time for values of the parameters in the weak segregation limit.
The linear behavior is clearly visible, characterized by
a constant position of the peak. It is followed by a sharp
transition to the asymptotic regime, during which the two plotted
quantities decay as
power laws. If $k_m(t)$ is fitted with $(t/\log t)^{1/z}$, the
computed exponent is $1/z = 0.253 \pm 0.003$,
in good agreement with the theoretical value $z=4$.

For deep quenches in the strong segregation limit,
the situation is quite different (Fig.~\ref{Fig3}).
At the end of the linear regime, the onset of the very late stage
dynamics is preceded by the two preasymptotic behaviors mentioned
in the previous section.
During the first one the system undergoes an almost complete stop;
therefore
$k_m(t)$ remains at the value of the linear regime, while $S_2(t)$
shows a plateau which extends over many decades.
This temporary stop in the evolution of the system is even better
illustrated by plotting directly the structure factor $C(k,t)$ for
different times during the pinning regime (Fig.~\ref{Fig3bis}).
These curves are compared with the analytic expression~(\ref{Cp})
of $C_p(k)$ in the pinning limit.
Later the dynamics restarts and is dominated by the time
dependent mobility.
The peak position and $S_2(t)$ go to zero as power-laws, in good
agreement with 
the expected behavior, $k_m \sim t^{-1/6}$ and $S_2 \sim t^{-1/3}$.
The agreement is not perfect because the system is already crossing over
to the asymptotic behavior. The onset of this last regime can
be delayed by making $\chi / \sigma^2$ bigger, but this would
also increase the duration of the pinned stage, making the
surface diffusion regime numerically unreachable.
Finally, on times longer than those shown in the figure, both quantities
smoothly cross over to the asymptotic behavior, which is the same of the
WSL.

The duration $\tau_p$ of the pinning is displayed in Fig.~\ref{Figtau}.
In the upper part $\tau_p$ is plotted versus $\chi/\sigma^2$
for $N$ strictly infinite, showing a power
law behavior whose measured exponent is
$0.49 \pm 0.01$, in very good agreement with the analytical estimate of
Eq.~(\ref{tau_power}).
In the lower part the same quantity, computed for $\sigma=0$,
is plotted versus $\chi N$, displaying an exponential
dependence as predicted in Eq.~(\ref{tau_exp}).

In the end, all figures confirm the analytical results discussed
above and the existence of a complex structure of intermediate
regimes and crossovers, as summarized in Table 1.

\section{Discussion}
\label{Sec5}

The solution of the large-$n$ model for phase-separating polymer blends
leads naturally to a comparison with the analogous results for small
molecule systems. In this way we can identify which of the modifications
introduced by the macromolecular nature of the blend components are
relevant.
We consider the effect of three modifications:
1) the mobility depending on the local order parameter and in particular
vanishing in pure phases;
2) the local potential having a double well form, but a logarithmic
expression, as opposed to the usual polynomial; 
3) the square gradient coefficient in the free energy having an
additional contribution
depending on the local order parameter, giving rise to two new terms in
the chemical potential.
The first two differences are actually not restricted to polymer blends
and can be considered also for small molecules; the third
is instead strictly related to the macromolecular nature of the
mixture.

In the WSL our results confirm what was already
known from previous numerical simulations of the full continuum
equation\cite{Chakrabarti89}.
The system belongs to the same universality class of small
molecule blends and all additional terms of the model turn out to be
irrelevant during the whole dynamical process.
This is not surprising when the scalar system is considered:
in the WSL separated phases are not pure: the equilibrium order
parameter in phase-separated domains is far from -1 and 1;
in such conditions $M(\psi)$ and $\kappa(\psi)$
have small local variations negligible compared to the 
average constant terms.

The situation is much more interesting in the SSL.
It turns out that all three modifications are relevant in this case.
The logarithmic expression for $f_{FH}$ pushes the minima of the local
potential close to -1 and +1 exponentially with $\chi N$.
For these limit values the mobility vanishes, and the evolution is pinned.
During the subsequent regime the non constant mobility is relevant since
the evolution is governed by $\Lambda(t)$ which owes its time dependence
to the order parameter dependence of $M(\psi)$.
Finally, as already pointed out in Sec.~\ref{SecLabel}, the asymptotic
stage is governed by the growing length $L(t)$ which is formally the same
of the small molecule case but is actually the result of the field
dependence of $M(\psi)$ and $\kappa(\psi)$.

Our aim is also to make statements about the real systems, not only
about the large-$n$ approximation to their temporal evolution.
Therefore a word must be said about the delicate
problem of the connection between the systems we want to study
(scalar order parameter) and those we are able to solve analytically
(vectorial order parameter with an infinite number of components).
For the small molecule case we already know that some properties of
the solution for large-$n$ model do not hold for scalar systems.
One of them is the multiscaling symmetry of the structure factor for
long times: for finite $n$ scaling symmetry holds.
Another difference is the value of the dynamical exponent $z$ which is known
to be 3 for scalar systems while is 4 when the order parameter is vectorial
(including the large-$n$ model).

These differences are known and can therefore be easily taken into account.
More dangerous may in principle be another difference between the
large-$n$ and the corresponding scalar model: the latter forms
ordered domains separated by well defined interfaces;
the former does not support interfaces and actually evolves towards
a state that is not truly ordered~\cite{Castellano97}.
This difference could be critical for a polymer blend in which interfaces
play a key role, through the field-dependence of the mobility and of
the square gradient coefficient.
Nevertheless we believe that in this case the picture provided by the
large-$n$ model is a close representation of what actually goes on
in scalar systems.

This conclusion relies on the comparison of our results with those
obtained for small molecule Ginzburg-Landau systems within the
large-$n$ approximation with constant and non constant mobility.
All the behaviors we find can be found also in large-$n$ models for
Ginzburg-Landau systems, where they are interpreted as the result of
different physical mechanisms governing growth.
By analogy we can describe our large-$n$ results as the effect of the
interplay of competing coarsening mechanisms for scalar polymer blends.
The pinning regime, the subsequent regime characterized by $z=6$ and
the asymptotic stage are all very clearly interpreted in terms of the
growth processes occurring in scalar polymer mixtures.

In particular, two are the mechanisms driving coarsening in binary blends.
The first is bulk diffusion, also said Lifshitz-Slyozov
or evaporation-condensation mechanism: A-molecules evaporate from
high curvature regions of A-rich domain interfaces; they diffuse in
B-rich regions and condensate on A-rich domains with lower curvature.
This process makes smaller domains shrink and larger grow; it is
associated with a $t^{1/3}$ growth law becoming $t^{1/4}$ for $n>1$.
The competing mechanism is the diffusion of molecules along domain
surfaces in order to minimize the interfacial energy.
This process has the effect of changing the shape (but not the volume)
of single domains and is associated with $z=4$ ($z=6$ in the corresponding
large-$n$ limit).
The slower growth law explains why surface diffusion is not observed in
usual small molecule systems and in polymers in the WSL: bulk
diffusion always prevails.
For deep quenches instead both mechanisms are slowed down, but in different
fashions.
Surface diffusion depends little on temperature and is only weakly
suppressed when $T \to 0$.
On the other hand the evaporation  process needed
for the Lifshitz-Slyozov mechanism is activated and therefore exponentially
inhibited for deep quenches: its probability is proportional
to $\exp(-\Delta F /k_BT)$ and the free energy change involved by the
evaporation of a macromolecule is $\Delta F \sim k_B T \chi N$.
In this way we can interpreted the succession of stages occurring in the
SSL: when both growth mechanisms are inhibited the system is pinned;
later growth starts, driven by surface diffusion, which is slow but only
weakly suppressed.
Eventually bulk diffusion prevails and phase-separation enters its late
stage.

Within this context it is difficult to understand why the pinning
limit requires $\chi/\sigma^2 \to \infty$ in addition to $\chi N \to
\infty$.
It is plausible that the additional condition is required only for the
large-$n$ model and does not apply to scalar systems. This is
suggested by the observation that the condition $\chi / \sigma^2
\to \infty$ is needed in order to neglect the term proportional to
$S_2(t)$ in Eq.~(\ref{eqC}).
Such term appears in the equation with same role of those derived from
the local potential in the free energy.
However it actually comes from the nonlocal part of the free energy:
it becomes ``local'' (i.e. not proportional to $k^2$) only as effect
of the large-$n$ limit.
It is very likely that in the scalar case the evolution freezes even if
$\chi /\sigma^2 < \infty$.

We finally discuss the relevance of the previous results in the
interpretation of the pinning phenomenon that has been observed
in off-critical quenches.
Experiments show that some polymeric mixtures quenched in the
unstable region of their phase diagram dramatically change their
behavior depending on the average concentration of the blend
components~\cite{Hashimoto92,Hashimoto94}.
When the concentration is critical growth proceeds as usual.
When concentration is sufficiently off-critical coarsening starts
but later stops, before the system reaches equilibrium,
in a frozen configuration with partially separated phases.
The specific mechanism responsible for this phenomenon is still poorly
understood and this topic has been the subject of discussion recently
\cite{Kotnis92,Castellano95}.
The conjecture that inhibition of bulk diffusion due to free energy
barriers may play a fundamental role has been put forward
\cite{Hashimoto92} but so far no convincing test of this hypothesis
has been done: direct numerical integration of the full equation of
motion is easily performed only in the WSL and no pinning has been
detected~\cite{Castellano95};
for deep quenches spurious numerical instabilities arise.
Using the large-$n$ limit approximation we are able to investigate the
strong segregation limit.
From our study a plausible explanation of the experimental evidence
comes out.
For extremely deep quenches all growth mechanisms are suppressed and
the system is pinned in a configuration out of equilibrium.
This is the pinning described in Sec.~\ref{SecPin} and does not
depend on concentration, i.e. it happens also for critical quenches.
It is very unlikely that this kind of pinning is observed in experiments,
since it probably requires unrealistically low temperatures.
The pinning phenomenon observed experimentally in instead more likely
related to intermediate values of $\chi N$, such that bulk diffusion
is inhibited while surface diffusion is not. This would explain
both the unarrested growth for critical quenches and the freezing for
off-critical ones.
When the concentration is critical an interconnected pattern is formed
for both phases and surface diffusion can drive the system to
macroscopical phase-separation.
When the quench is sufficiently off-critical instead, the minority
phase forms non percolating droplets embedded in a matrix of the
majority phase.
Surface diffusion can only lead to a partial phase-separation and
coarsening stops when droplets are spherical.
However, on much longer times, the residual bulk diffusion should
drive the system to complete phase-separation.
In order to confirm this scenario further work on the numerical
solution of the scalar order parameter equation in the SSL is in
progress.
More experiments, aimed at verifying the prediction that coarsening
should restart for very long times after the pinning in off-critical
quenches, would also be very helpful.

We thank Marco Zannetti for an interesting discussion.

\section*{Appendix 1}
We consider the Flory-Huggins-De Gennes free energy functional $F(\psi)$ 
and the mobility $M(\psi)$; in order to generalize them to the vector 
order parameter case we require 
$F_V({\vec \psi} )$ and $M_V({\vec \psi} )$, the vectorial counterparts 
of the free energy and of the mobility, to be ${\cal O}(n)$ symmetric 
functions of 
${\vec \psi} ({\bf x},t)$. With this position the field dependence occurs
through the modulus of vector quantities, namely $|{\vec \psi}
({\bf x},t)|= \left[\sum _{\beta=1}^{n}\psi_\beta^2
({\bf x},t)\right]^{1/2}$
and $|{\bf \nabla} {\vec \psi} ({\bf x},t) |=
\left\{\sum _{\beta =1}^{n} [\nabla _\beta \psi
_\alpha ({\bf x},t)]^2\right \}^{1/2}$.
In the large-$n$ limit if one requires the 
single component $\psi _\alpha ({\bf x},t)$ to remain finite, the
square modulus 
of vector quantities must be normalized by $1/n$, in order to keep it
finite.
Hence the whole field dependence of $F_V({\vec \psi} )$ and 
$M_V({\vec \psi} )$ occurs in the vectorial case through 
$|{\bf \overline \psi}({\bf x},t)|$
and $|{\bf \nabla} {\bf \overline \psi} ({\bf x},t) |$, where 
${\bf \overline \psi}({\bf x},t)=n^{-1/2}{\vec \psi}({\bf x},t)$.
Moreover
one requires $F_V({\vec \psi} )$ to be an extensive quantity in the number 
of components $n$. In summary, a proper generalization to the vector
case is achieved by substituting everywhere $\psi({\bf x},t)$ and 
$|{\bf \nabla} \psi ({\bf x},t) |$ with $|{\bf \overline \psi}({\bf x},t)$
and $|{\bf \nabla} {\bf \overline \psi} ({\bf x},t) |$ respectively in 
Eqs.~(\ref{ham}), ~(\ref{kappa}), ~(\ref{localpot}),
~(\ref{mobility}), and multiplying $F({\bf \psi} )$ by $n$.
We obtain 
\begin{equation}
\frac{F_V({\vec \psi } )}{k_BT}=n \int d{\bf x} 
\left \{ \frac{f_{FH}(|{\bf \overline \psi }|)}{k_BT}+
\kappa(|{\bf \overline \psi} |)
|\nabla {\bf \overline \psi } ({\bf x},t)|^2 \right \}
\end{equation}
where $\kappa(|{\bf \overline \psi} |)$ and
$f_{FH}(|{\bf \overline \psi }|)$
are still given by expressions~(\ref{kappa}) and~(\ref{localpot}).
The Cahn-Hilliard equation for the time evolution of a generic component 
$\alpha$ of the vector field ${\vec \psi } ({\bf x},t)$ reads
\begin{equation}
\frac {\partial \psi _\alpha ({\bf x},t)}{\partial t}={\bf \nabla} \cdot \left [
M_V({\vec \psi }){\bf \nabla} \frac{\delta F_V({\vec \psi })}
{\delta \psi_\alpha}\right ]
\end{equation}
where $M_V({\vec \psi })=M(|{\bf \overline \psi}|)$ is given by 
Eq.~(\ref{mobility}) and we have neglected thermal noise as discussed 
in Sec.~\ref{Sec2}.
Then, considering a symmetric blend, the Langevin evolution
of $\psi _\alpha ({\bf x},t)$ is obtained 
\bea \nonumber
\frac{\partial \psi _\alpha ({\bf x},t)}{\partial t}&
\!\!=\!\!&\frac{ND}{2} \!{\bf \nabla} \cdot \!
\left \{\left(1-|{\bf \overline \psi} ({\bf x},t)|^2\right)
\!{\bf \nabla}\! \left [\frac{1}{4N}\ln  \left ( \frac 
{1+|{\bf \overline \psi} ({\bf x},t)|}
{1-|{\bf \overline \psi} ({\bf x},t)|}
\right )\frac{\psi _\alpha ({\bf x},t)}{|{\bf \overline \psi}
({\bf x},t)|} + \right. \right. \\ \nonumber
& & -\frac{\chi }{4} \psi _\alpha ({\bf x},t)
-\left (\lambda ^2 \chi+\frac {\sigma ^2}
{9(1-|{\bf \overline \psi} ({\bf x},t)|^2)}
\right ){\bf \nabla} ^2 \psi _\alpha ({\bf x},t) + \\ 
& & \left. \left. -\frac{\sigma ^2 \overline \psi
_\alpha ({\bf x},t)}{9(1-|{\bf \overline \psi} ({\bf x},t)| ^2)^2}
|{\bf \nabla} \overline \psi _\alpha ({\bf x},t) |^2\right ] \right \}
\label{n=n}
\eea
For $n=1$ one recovers Eq.~(\ref{n=1}). In the large-$n$ limit,
summing over vector components
averages the system over an ensemble of configurations, and hence
\begin{equation}
\lim _{n\to \infty}|{\bf \overline \psi} ({\bf x},t)|^2=
\lim _{n\to \infty}\frac{1}{n}\sum _{\beta =1}^{n}\psi _\beta ^2({\bf x},t)
=<\psi_{\alpha}^2({\bf x},t)>\equiv S(t)
\end{equation}
where $<. . . >$ denotes the ensemble average, translational
invariance has been assumed and $S(t)$ does not depend on $\alpha$
due to internal symmetry.
Analogously
\begin{equation}
\lim _{n\to \infty}|{\bf \nabla} \overline \psi _\alpha ({\bf x},t)|^2=
\lim _{n\to \infty}\frac{1}{n}\sum _{\beta =1}^{n} [\nabla _\beta \psi
_\alpha ({\bf x},t)]^2=<[\nabla \psi_\alpha ({\bf x},t)]^2>\equiv S_2(t)
\end{equation}
Hence Fourier transforming to reciprocal space, the evolution equation for
the order parameter field reads
\bea \nonumber
\frac{\partial \psi ({\bf k},t)}{\partial t}&=&-\frac{ND}{2}[1-S(t)]k^2
\left \{\frac{1}{4NS^{\frac{1}{2}}(t)}\ln \left( 
 \frac 
{1+S^{\frac{1}{2}}(t)}{1-S^{\frac{1}{2}}(t)} \right)
-\frac{\chi }{4} + \right. \\
& & \left. 
+\frac{\sigma ^2}{9}\frac{S_2(t)}{[1-S(t)]^2}
+\left (\lambda ^2
\chi+\frac {\sigma ^2}{9[1-S(t)]}\right )k^2 
\right \}\psi ({\bf k},t)
\eea
where the component index $\alpha$ has been dropped.

\section*{Appendix 2}
In this Appendix we derive the expressions~(\ref{tau_exp}) and
(\ref{tau_power})
for the duration of the pinned stage when the pinning limit is approached.
We define $\tau_p$ with reference to the behavior of the quantity $S_2(t)$.
As can be seen in Fig.~\ref{Fig3} $S_2(t)$ displays a plateau
during the pinned stage.
More precisely it reaches a maximum for $t=t_p$ at the end of the linear
regime and decreases extremely slowly until the crossover
time $t=t_p+\theta_p$, when the pinned stage ends and the system enters
the subsequent time evolution characterized by a more rapid (power law)
decrease of $S_2(t)$.
A quantitative definition of $\theta_p$ can be obtained from the relative
variation of $S_2$ by letting
\begin{equation}
{S_2(t_p)-S_2(t_p+\theta_p) \over S_2(t_p)} = \epsilon
\label{defthetap}
\end{equation}
where $\epsilon$ is an arbitrarily fixed small number.
Since $S_2(t)$ is approximately constant during the pinned stage one has
\begin{equation}
S_2(t_p+\theta_p) = S_2(t_p) + 
\left. \frac{\partial S_2(t)}{\partial t} \right|_{t=t_p} \theta_p
\end{equation}
and therefore, using Eq.~(\ref{defthetap})
\begin{equation}
\theta_p = - \epsilon S_2(t_p) \left[
\left. \frac{\partial S_2(t)}{\partial t} \right|_{t=t_p}  \right]^{-1}
\label{A1}
\end{equation}
$\theta_p$ is the actual duration of the pinned regime
but it should be noticed that in the pinning limit $\partial C/\partial t$
is proportional to $\chi N$, which goes to infinity.
All times are divided by this factor and hence vanish.
In order to compare the duration of the pinned stage for different
values of $\chi N$, $\theta_p$ must be rescaled by the appropriate intrinsic
time factor $1/\chi N$; we therefore define the duration $\tau_p$ of
pinning as
\begin{equation}
\tau_p = \chi N \theta_p
\label{A3}
\end{equation}
The derivative in Eq.~(\ref{A1})
can be computed by considering that $C(k,t=t_p) \simeq C_p(k)$
defined in Eq.~(\ref{Cp}) so that
\begin{equation}
\left. \frac{\partial S_2(t)}{\partial t} \right|_{t=t_p} =
\int \frac{d^d{\bf k}}{(2\pi)^d} k^2
\left. \frac{\partial C ({\bf k},t)}{\partial t} \right|_{t=t_p} 
\simeq - \chi N D (1-S_p) (\lambda^2 S_{6p} - S_{4p}/4)
\label{A2}
\end{equation}
where 
\begin{equation}
S_{np} = S_n(t_p) = \int \frac{d^d{\bf k}}{(2\pi)^d} k^n C_p(k)
\end{equation}
are known quantities.
By inserting Eqs.~(\ref{A1}) and~(\ref{A2}) in
Eq.~(\ref{A3}) one obtains an expression
for $\tau_p$
\begin{equation}
\tau_p \simeq {\epsilon S_{2p} \over 
D (1-S_p) (\lambda^2 S_{6p} - S_{4p}/4)}
\end{equation}
where only the value $S_p$ of $S(t)$ during the pinned stage
remains to be determined.
This is calculated by imposing that
\begin{equation}
\left. \frac{\partial S(t)}{\partial t} \right|_{t=t_p} = 0
\label{Sp}
\end{equation}
When evaluating this condition the outcome depends on how the pinning
is approached, i. e. on the order of the limits $\chi N \to \infty$ and
$\chi/\sigma^2 \to \infty$.
When one takes $\chi/\sigma^2 \to \infty$ with large but fixed $N$, one has
\begin{equation}
0 = - \chi N D (1-S_p) \left\{ \left[ -1/4 + {1 \over 4 \chi N} \ln \left(
{2 \over 1 - S_p^{1/2}} \right) \right ] S_{2p} + \lambda^2 S_{4p}
\right \}
\end{equation}
With simple algebra one obtains
\begin{equation}
1-S_p = 4 \exp \left\{ -\chi N \left[ (S_{2p}-4\lambda^2 S_{4p})
\over S_{2p} \right] \right\}
\end{equation}
and therefore
\begin{equation}
\tau_p \sim \exp \left\{ \chi N \left[ (S_{2p}-4\lambda^2 S_{4p})
\over S_{2p} \right] \right\}
\end{equation}
Letting instead $\chi N \to \infty$ with non vanishing $\sigma^2$
\begin{equation}
0 = - \chi N D (1-S_p) \left\{
\left[ -1/4 + {\sigma^2 \over \chi}
{S_{2p} \over 9 (1-S_p)^2} \right] S_{2p}
+ \left[ \lambda^2 + {\sigma^2 \over \chi} {1 \over 9(1-S_p)} \right]
S_{4p} \right \}
\end{equation}
yielding
\begin{equation}
1-S_p = \sqrt{S_{2p}^2 \over 9(S_{2p}/4-\lambda^2 S_{4p})}
\left({\chi \over \sigma^2} \right)^{-1/2}
\end{equation}
and therefore
\begin{equation}
\tau_p \sim \left({\chi \over \sigma^2} \right)^{1/2}
\end{equation}

\section*{Figures}

\begin{figure}
\begin{center}
\def\epsfsize#1#2{0.65\textwidth}
\leavevmode
\epsffile{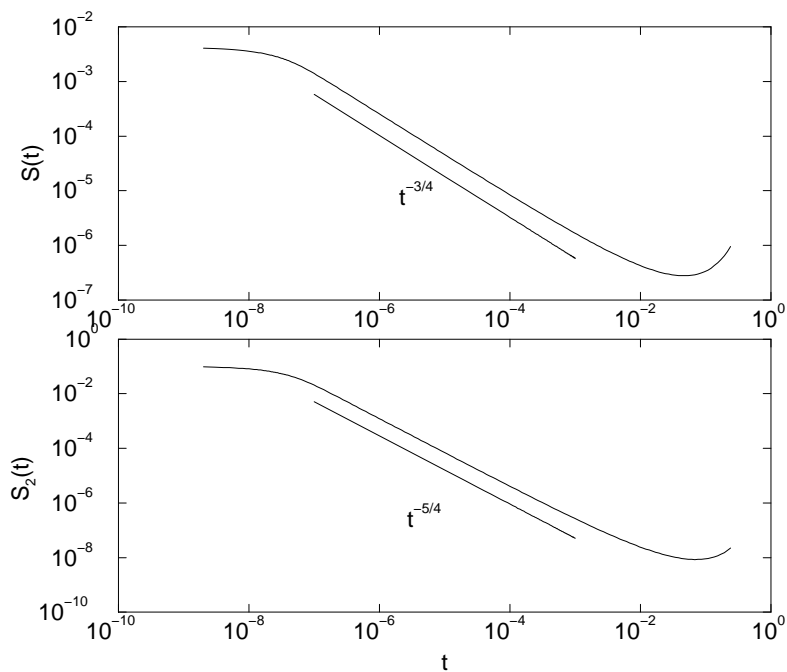}
\end{center}
\caption{
\label{Fig1}
Top: Log-log plot of $S(t)$ vs $t$ for early times. The values of
the parameters are $N=0.25 \times 10^5$, $\chi N=10^3$
$\lambda=1/2$, $\sigma=1$.
Bottom: The same plot for $S_2(t)$.
}
\end{figure}

\begin{figure}
\begin{center}
\def\epsfsize#1#2{0.65\textwidth}
\leavevmode
\epsffile{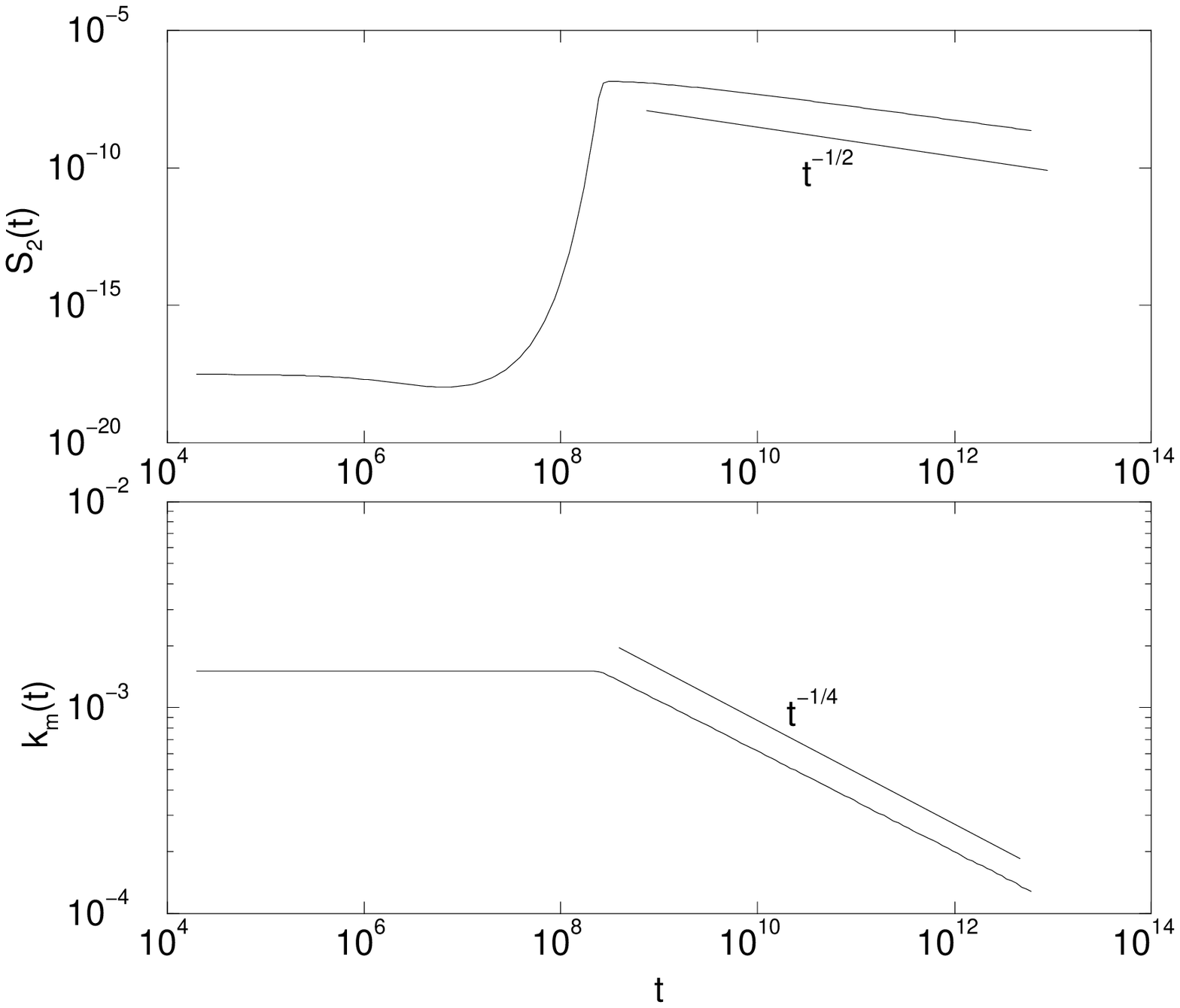}
\end{center}
\caption{
\label{Fig2}
Top: Log-log plot of $S_2(t)$ vs $t$ for a quench in the weak
segregation limit. The values of the parameters are
$N=0.25 \times 10^5$, $\chi N=2.1$, $\lambda=1/2$, $\sigma=1$.
Bottom: The same plot for $k_m(t)$.
}
\end{figure}

\begin{figure}
\begin{center}
\def\epsfsize#1#2{0.65\textwidth}
\leavevmode
\epsffile{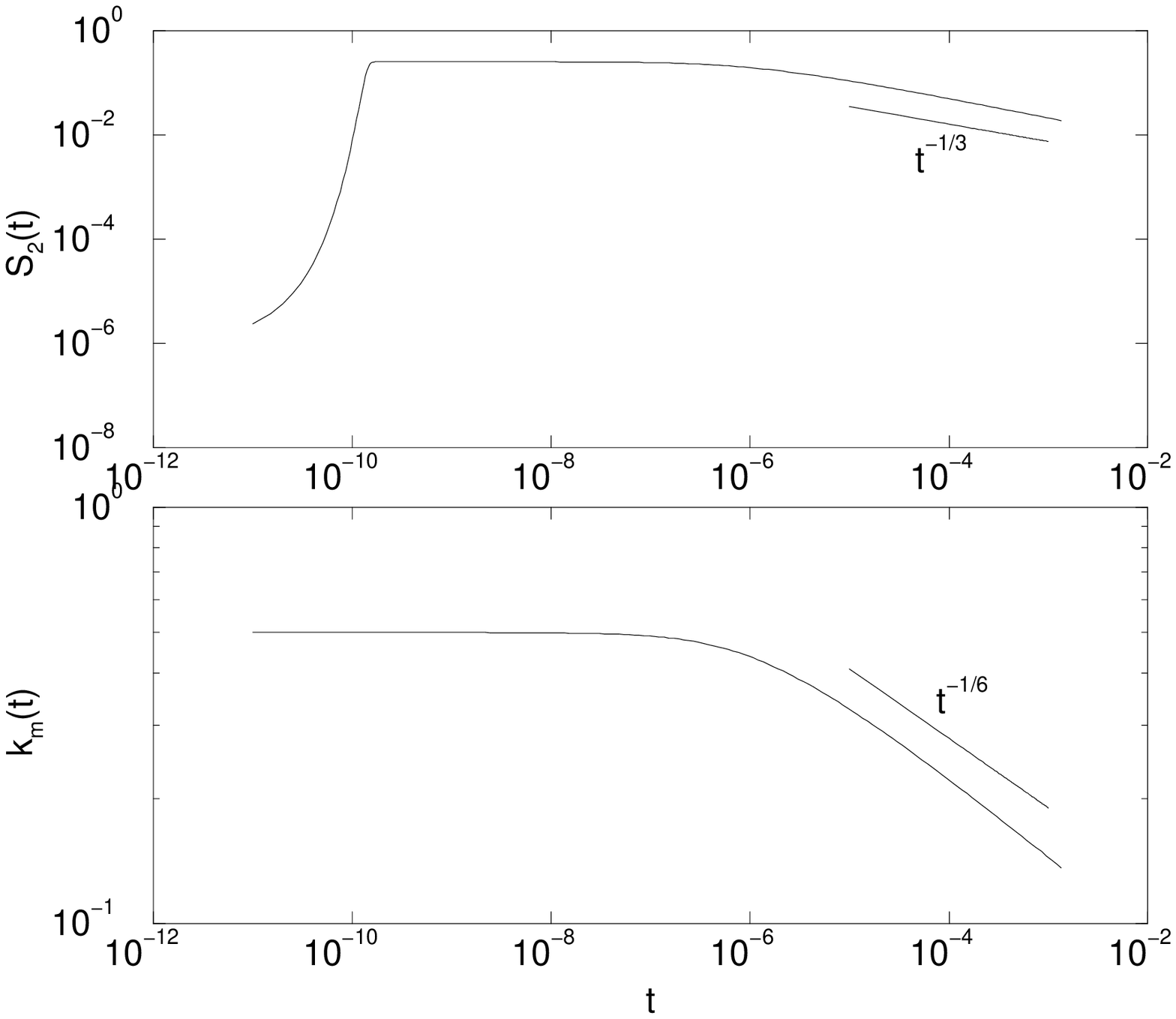}
\end{center}
\caption{
Top: Log-log plot of $S_2(t)$ vs $t$ for a quench in the strong
segregation limit. The values of the parameters are
$N=0.25 \times 10^5$, $\chi N = 10^{12}$, $\lambda=1/2$, $\sigma=1$.
Bottom: The same plot for $k_m(t)$.
\label{Fig3}
}
\end{figure}

\begin{figure}
\begin{center}
\def\epsfsize#1#2{0.65\textwidth}
\leavevmode
\epsffile{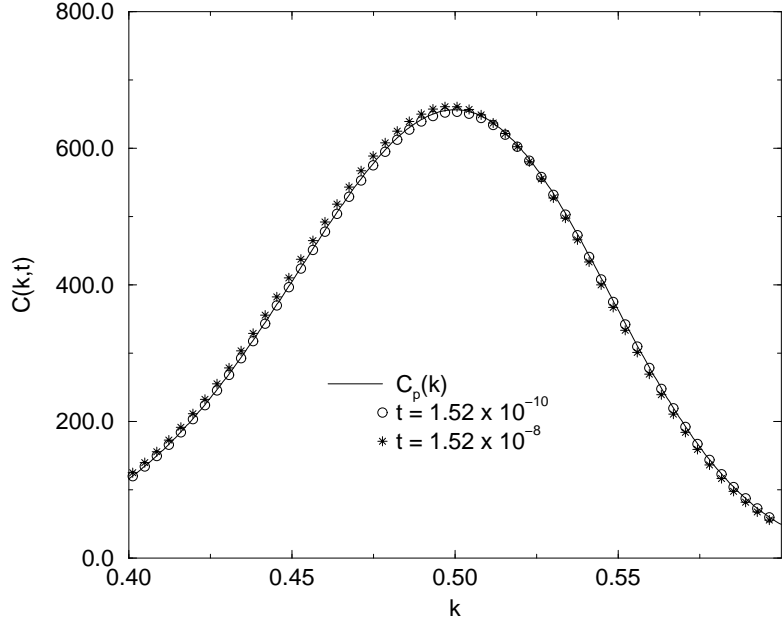}
\end{center}
\caption{
Plot of $C(k,t)$ vs $t$ for the same parameters of
Fig.~(\ref{Fig3}) and two different times separated by two decades,
compared with the analytical expression (Eq.~(\ref{Cp})).
\label{Fig3bis}
}
\end{figure}

\begin{figure}
\begin{center}
\def\epsfsize#1#2{0.7\textwidth}
\leavevmode
\epsffile{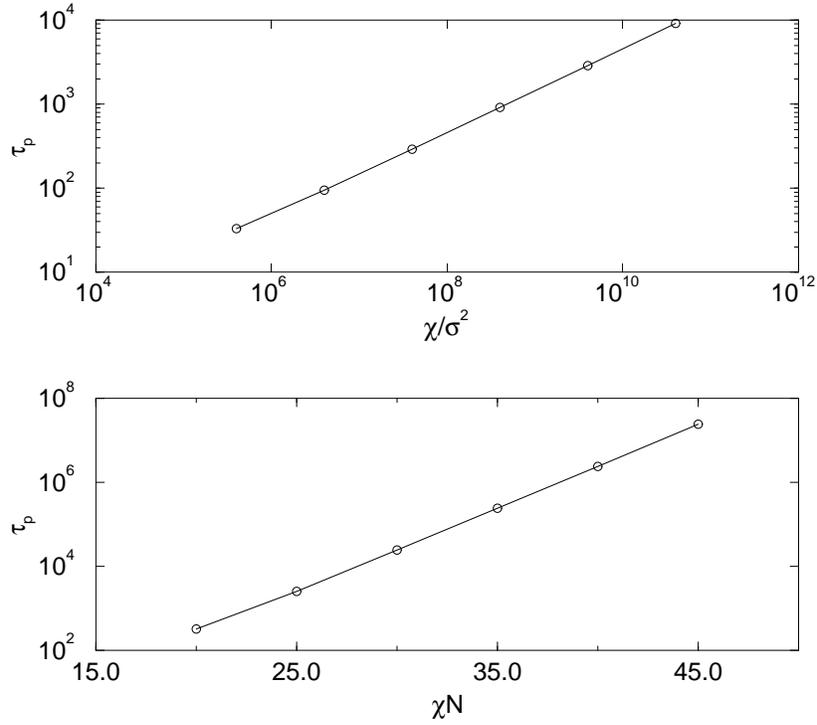}
\end{center}
\caption{
Top: Log-log plot of $\tau_p$ vs $\chi/ \sigma^2$ for $N = \infty$
showing the power law divergence of the pinning duration.
Bottom: Linear-log plot of $\tau_p$ vs $\chi N$ for $\sigma = 0$,
displaying that in this limit $\tau_p$ diverges exponentially.
In all cases $\tau_p$ was determined by choosing $\epsilon=10^{-4}$.
\label{Figtau}
}
\end{figure}

\section*{Table}
\begin{center}
\begin{tabular}{||l|c|c|c|c|c||} \hline
Regime & Early scaling & Linear & Pinned & Surface Diff. & Asymptotic \\ \hline
$S(t)$ & $t^{-d/4}$ & $\exp(t)$ & const & const & const \\
$S_2(t)$ & $t^{-(d+2)/4}$ & $\exp(t)$ & const & $t^{-1/3} (\log t)^{-1/2}$&
$t^{-1/2} (\log t)^{-1/2}$ \\
$k_m^{-1}(t)$ & const & const & const & $t^{1/6}(\log t)^{-1/4}$&
$(t/\log t)^{1/4}$ \\
${\cal L}(t)$ & $t^{1/2}$ & $t^{1/2}$ & const & $t^{1/6}(\log t)^{1/4}$&
$(t \log t)^{1/4}$ \\
$\Lambda(t)$ & $t^{1/4}$ & $t^{1/4}$ & const & $t^{1/6}$&$t^{1/4}$ \\
\hline
\end{tabular}
\end{center}

\noindent
Table 1. Summary of the time dependence of the important quantities
during the different stages.
The third and fourth time regimes are observable only when $\chi N \gg 1$ and 
$\chi / \sigma^2 \gg 1$.

\end{document}